\documentclass[11pt]{article}
\usepackage{mwe}
\usepackage{wrapfig}
\usepackage{amssymb}
\usepackage{xcolor}
\usepackage{mathtools}
\usepackage{soul,xcolor}
\usepackage{ulem}
\usepackage{graphicx}
\usepackage{authblk}

\newcommand{\red}{\textcolor{red}}
\newcommand{\blue}{\textcolor{blue}}

\newcommand{\ind}{$\mbox{ }$\hspace{2mm}}

\newcommand{\vs}{\vspace{2mm}}

\newcommand{\dind}{\ind \ind}

\newcommand{\fig}{\includegraphics}
\newcommand{\ahmm}{$\alpha$-HMM}
\newtheorem{definition}{Definition}

\newcommand{\bX}{{\bf X}}
\newcommand{\bY}{{\bf Y}}
\newcommand{\lb}{\langle}
\newcommand{\rb}{\rangle}
\newcommand{\ptr}{^{(3)}}
\newcommand{\pfv}{^{(5)}}

\begin{document}

\title{$\alpha$-HMM: 
A Graphical Model for RNA Folding}

\author[1]{\sc Sixiang Zhang\footnote{These authors contributed equally to this work. $^\dagger$Correspondence  email: {\tt liming@uga.edu}}}
\author[2]{\sc Aaron J. Yang$^*$}

\author[1]{\sc Liming Cai$^{\dagger}$}

\affil[1]{\vspace{2mm}School of Computing, University of Georgia, Athens, GA 30602}

\affil[2]{North Oconee High School,
Bogart, GA}

\date{December 15, 2023}
\maketitle
\begin{abstract}

\noindent
RNA secondary structure is modeled with the novel arbitrary-order hidden Markov model ($\alpha$-HMM). The $\alpha$-HMM extends over the traditional HMM with capability to model stochastic events that may be influenced by historically distant ones, making it suitable to account for long-range canonical base pairings between nucleotides that constitute the RNA secondary structure. Unlike previous heavy-weight extensions over HMM, the $\alpha$-HMM has the flexibility to apply restrictions on how one event may influence another in stochastic processes, enabling efficient prediction of RNA secondary structure  including pseudoknots.
\end{abstract}
\section{Introduction}

The secondary structure of a ribonucleic acid (RNA) is higher order structure over 
the primary sequence of the molecule. Nucleotides on the sequence physically come close to each other through hydrogen bonds between bases, forming canonical Watson-Crick pairs ({\tt A-U} and {\tt G-C}), and the wobble pair ({\tt G-U}) as the fundamental components of the structure. The secondary structure is the intermediate,  to a great extent the scaffold for higher order interactions between nucleotides to generate RNA tertiary, i.e., 3-dimensional, structure 
\cite{DoudnaEtAl1998,TinocoAndBustamante1999}.
The latter determines important RNA functions in  biological processes, not only as a genetic information carrier but also playing catalytic, scaffolding, structural, and regulatory roles \cite{GearyEtAl2014,CechAndSteitz2014}. There has been abundant interest in  understanding the detailed process and dynamics of how RNA folds into its structure \cite{BateyEtAl1999}. Computational prediction of RNA secondary structure directly from its primary sequence is a very desirable step toward the prediction of RNA 3D structure. This is evident by the RNA Puzzles, an annual competition to predict RNA 3D structures, in which most of the used methods by participants are proceeded by a phase for secondary structure prediction \cite{MiaoEtAl2015,MiaoEtAl2017,MiaoEtAl2020}. 

In the past few decades, various secondary structure prediction methods have been developed based on a few different computational paradigms. Many of them have been built upon the energy minimization principle for the reason that canonical base pairs reduce free energy and increase the stability  of the folded structure. Methods based on this folding mechanism have evolved from the rudimentary maximizing counts of base pairs \cite{NussinovAndJacobson1980} to rather sophisticated energy calculations on various structure elements, arriving at prediction accuracy as high as 80\% \cite{Zuker2003,GruberEtAl2008,FallmannaEtAl2017} on short to moderately long RNA sequences. However, evidence has suggested that the energy minimization folding mechanism has to overcome huge obstacles in prediction from longer RNA molecules \cite{AmmanEtAl2013}. On the other hand, more recent developments in machine learning based prediction have shown that, with enough training data, they can outperform traditional folding mechanism-dependent methods, e.g., with accuracy exceeding the 90\% threshold \cite{ZhaoEtAl2021}. Unfortunately, the ML-based method are data-driven without 
understanding of the RNA folding mechanism. Moreover, with such a method the high performance of prediction has yet to see the generalization across different categories of secondary structures \cite{MathewsEtAl2022}. Until research in explainable deep learning \cite{RasEtAl2021} becomes mature and available, it is extremely difficult to deduce from ML-based methods a plausible accurate fold mechanism for advancement in our understanding RNA biology.

Statistical models have played important roles in RNA secondary structure folding. 
One typical model is the stochastic context-free grammar (SCFG) that has been effective for RNA secondary structure modeling and prediction. Based on Chomsky rewriting rules \cite{LariAndYoung1990}, an SCFG defines sentences of an interested language (e.g., sequences of RNAs) and every syntactic generation of an RNA sequence offers a potential structural interpretation for the sequence \cite{SakakibaraEtAl1994,EddyAndDurbin1994}. Specifically, every pair of long-range, covariant nucleotide bases are generated at the same time by the rewriting rules. Association of probabilities with grammar rules probabilistically  defines a structure ensemble for every RNA sequence. SCFG models have been shown to be equivalent to the energy minimization based methods for secondary structure prediction. Therefore, likewise, SCFG also fails to model RNA pseudoknots, a type of secondary structure with crossing patterns (i.e., context-sensitive) of base pairs.
Coping with context-sensitivity in computational linguistics has yielded some hybrid probabilistic systems \cite{JoshiAndRambow2003}, where decoding optimality and computation efficiency cannot be both prioritized.  
It has been a challenging quest to discover  
statistical models that can rigorously characterize higher order structures over RNA sequences while ensuring efficiency in optimal structure prediction.

In this paper, we introduce the arbitrary-order hidden Markov model (\ahmm) for modeling and prediction of RNA secondary structure including pseudoknots. Decoding hidden information from sequential data through hidden Markov models has been quite a success 
\cite{Eddy1996Hidden,Eddy1998Profile,Rabiner1986Introduction,Koski2001Hidden,Smyth1997Probabilistic}. An HMM is represented by a probabilistic graph of state transitions; 
a random walk on states  is a stochastic process, generating a Markov chain of events with the joint probability simply factored into conditional probabilities between two consecutive events. To account for higher order structure over finite stochastic processes, in an $\alpha$-HMM, the probabilistic graph of state transitions is equipped with additional edges (called {\it influences}) than can relate historically distant events to recent ones. Technically, the influence from one event to another is based on the ``nearest influence principle'', making the \ahmm\, very suitable for modeling crossing as well as nested and parallel base pairs. 
The $\alpha$-HMM is different from previous works on extensions of the HMM, typically the $k^{\rm th}$-order HMMs, where influencing historical events are all within a given distance $k$, for some fixed $k\geq 1$, to the current event
\cite{SerranoAlfaro2017Limitations,Maruvada2017Thesis,Koski2001Hidden,Zaki2010Vogue}. 
Unlike previous heavy-weight extensions over HMM, the $\alpha$-HMM has the flexibility to apply restrictions on how one event may influence another in stochastic processes. 


With the \ahmm, (arbitrary) long-range canonical base pairings within an RNA sequence are modeled as 
stochastic events that are influenced by distant, historical ones. 
Specifically, a base pairing is permitted when the stochastic process enters some special state that can be influenced by another (historical) state. Every influence is modeled with a non-directed edge in the probabilistic graph. Semantically, the influence is a coordination between paired nucleotide symbols emitted by the two states, which abide by certain probability distribution. We present a dynamic programming algorithm that maximizes the joint probability of the input RNA sequence and decodes the corresponding hidden states as the (most likely) secondary structure. In addition, the \ahmm-based RNA secondary structure prediction, is able to predict pseudoknotted structures. Unlike the SCFG, which limits nucleotide base pairings to the nested and parallel patterns, the \ahmm\, permits two crossing influence edges, making it possible to predict crossing patterns of nucleotide base pairs, i.e., pseudoknots.  The algorithm is appealing also because it runs in $O(n^3)$ on the input RNA sequence of length $n$, much more efficient than the state-of-the-art algorithms for RNA pseudoknot prediction \cite{JabbariEtAl2018,SatoAndKato2022}.

\section{Prelimiaries}

\subsection{Probabilistic influence graph}

We use $[n]$ to denote set $\{1,2,\dots,n\}$ and $[0..1]$ for the closed real interval between 0 and 1. 

\vspace{-3mm}
\begin{definition} 
\rm Let $\Sigma$ be a finite alphabet and $G=(S, T\cup A)$ be a finite, directed multi-graph\footnote{A multi-graph is a graph that may contain multiple edges between two vertices and self-loops (edges from one vertex to itself).} over 
vertex (state) set $S$, where $T$ and $A$ are 
two directed edge sets. Let $\epsilon, \tau$, and $\eta$ be probabilistic functions that   satisfy

\noindent
\dind 
(1) $\epsilon: S\times \Sigma \rightarrow [0,1]$, such that $\forall\, q\in S$, $\sum\limits_{a \,\in\, \Sigma} \epsilon (q, a) =1 $; \\
\dind (2) $\tau: T \rightarrow [0,1]$, such that $\forall\, q\in S$, $\sum\limits_{(q, r)\,\in \,T} \tau (q, r) =1$; \\
\dind (3) $\eta: A\times \Sigma^2 \rightarrow [0,1]$, such that $\forall\, (q, r) \in A$, 
$\sum\limits_{a, \,b \,\in \,\Sigma} \eta(q, r, a, b) =1$;\\
\dind (4) $\forall q \in S$, $(q, r) \in A$ and $(q, s) \in A$ imply $r=s$ unless $(q, q)\in T$.

\noindent
\dind (5) $\forall  q, r, s \in S$, either $(q, r) \not \in A$, or $(r, s) \not \in A$, unless $(r, r) \in T$.

\vspace{1mm}
The pair $\langle G, \Theta\rangle$, where $\Theta=\{\epsilon, \tau, \eta\}$, is called a {\it probabilistic influence graph} (PIG) 
over the alphabet $\Sigma$. $\epsilon, \tau$, and $\eta$ are called {\it emission, transition}, and {\it influence  functions}, {respectively}.
\end{definition}
\vspace{-3mm}

Condition (3) is about the extra probability 
contribution as the result of state $q$ influencing state $r$, 
where function $\eta$ measures the probability in terms of the relationship between the two states and their emitted symbols. Condition (4) restricts  that any state $q$ can only influence at most one other state unless
state $q$ is ``replicable'', i.e., with a self-loop transition edge. Condition (5) restricts that 
any influenced state $r$ cannot influence another state, 
 unless state $r$ is ``replicable''. We will show that these restrictions do not weaken the expressibility of PIGs.
\begin{definition}\rm
In a PIG, state $r\in S$ is {\it affiliated} with another state $q\in S$ if there is a transition edge $(q, r) \in T$ unless also $(r, r) \in T$. In addition, by default, state $r$ is affiliated with itself if $(r, r) \in T$.
\end{definition}
\begin{definition}\rm Let $\langle G, \Theta\rangle$ be a PIG and Let $n\geq 1$. A {\it walk} ({\it of $n$ steps}) on the PIG is a sequence of pairs $\rho =\lb q_1, k_1\rb \lb q_2, k_2\rb \dots \lb q_n, k_n\rb$, provided that  

\noindent
\dind (i) for $1\leq i\leq n$, 
$q_i \in S$, $k_i=0$, or $i<k_i \leq n$ and $(q_i, q_{k_i}) \in A$; and \\
\dind (ii) for any $i\geq 1$, if $k_i = j \not = 0$, for some $j$,  and $q_{j}$ is affiliated with $q_{j-1}$, then $k_{i+1} = j-1$.
\end{definition}
The first component of each pair $\lb q_i, k_i\rb$ is the state $q_i$ at step $i$; the second component $k_i$ is the step on which the state $q_i$ has an {\it impact} via an influence edge. Alternatively, $k_i=0$ simply means that state $q_i$ does not have an impact on any step  (regardless whether the influence edge $(q_i, r)$, for some state $r$, exists or not in the PIG). Condition (ii) assumes that, given two affiliated states $q_{j-1}$ and $q_j$ on the walk, if 
$q_i$ has impact on $q_j$ via some influence, then $q_{i+1}$ should have impact on $q_{j-1}$ via some influence. 

Figure~\ref{pig}(a) shows a small computational linguistic model in the PIG where states are assumed to emit words or phrases so walks on the PIG are expected to generate sentences. Dotted, colored arrows are influence edges to constrain relationships of emitted words or phrases between two related states. For example, the tense of a {\tt verb} may be influenced by a {\tt time} adverb. 

\begin{figure}[ht]
\begin{center}
\fig[width=180pt]{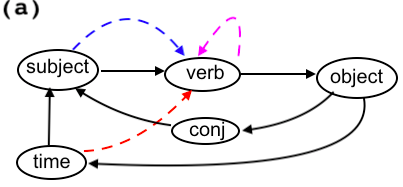}\,
\fig[width=220pt]{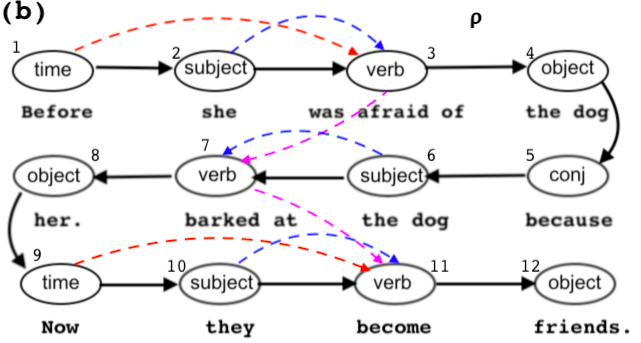}
 \end{center}
\caption{A schematic illustration of a PIG modeling a simplified subset of English sentences. (a) States are within oval circles which can emit desirable words and phrases (emissions are not shown); solid arrows are transitions, and dotted (colored) arrows are influences.  (b) A walk $\rho$ of length 12 starting from state {\tt time} to generate the observed sentence, with influences labeled among the instantiated states on the walk $\rho$.}
\label{pig} 
\end{figure}

\subsection{The \ahmm}
\begin{definition}\rm  Let $\rho = (q_1, k_1)  \dots (q_n, k_n)$ be any walk of length $n$ on a PIG. The {\it influencers} on steps of $\rho$  is a function  $F_{\rho}: [n] \rightarrow 2^{[n]}$ such that, for any $j$, $1\leq j \leq n$,
\[ F_{\rho}(j) = \{ i : \, k_i = j\} \cup \{j-1\}\]
\end{definition}

Let $\langle G, \Theta\rangle$ be a PIG as defined above. We are interested in the following two discrete stochastic processes {\it associated with} $\langle G, \Theta\rangle$: ${\bf X} =\{X_n: n\geq 1\}$ and $\bY =\{Y_n: n\geq 1\}$, where random variable 
$X_i$ and $Y_i$, $i\in [n]$, draws values from domain $S\times {\cal N}$ and $\Sigma$, respectively, where $\cal N$ is the set of natural numbers. For any $n\geq 1$, the joint probability of the first $n$ random variables in the stochastic processes can be computed with  using the chain rule.

\begin{equation}
\label{chainrule}
 P\big(\bigcup_{i=1}^n (X_i, Y_i) \big) = P\big ( X_1, Y_1\big) \prod_{i=1}^{n-1} P\big(X_{i+1}, Y_{i+1} \big | \bigcup_{j=1}^i (X_j, Y_j) \big )
 \end{equation}
 



\definition \rm
Let $\langle G, \Theta\rangle$ be a PIG as defined above and ${\bf X}$ and 
 $\bY$ be two associated discrete stochastic processed. Then 
the pair $(\bX, \bY)$ is an {\it arbitrary-order hidden Markov model} ($\alpha$-HMM) if, for $n\geq 1$,  on any walk $\rho = \lb q_1, k_1\rb \dots \lb q_n, k_n\rb $ and any sequence of symbols $a_1a_2 \dots a_{n}\in \Sigma^n$,
\begin{equation}
\label{ahmm}
\begin{split}
P &\big(X_n =\lb q_n, k_n\rb, Y_n =a_n\, \big |\, \bigcup_{i<n}  \big (X_{i}=\lb q_{i}, k_i\rb, Y_{i}=a_{i}\big )\big) \\
& = P\big(X_n =\lb q_n, k_n\rb, Y_n =a_n \,\big|\, \bigcup_{i \, \in \,F_\rho(n)}  \big (X_{i}=\lb q_{i}, k_i\rb, Y_{i}=a_{i}\big )\big )  \\
\end{split}
\end{equation}
where individual conditional probability is defined as, for $i<n$, 
\begin{equation}
\small
P\big(X_n =\lb q_n, k_n\rb, Y_n =a_n\, \big |\, X_{i}=\lb q_{i}, k_i\rb, Y_{i}=a_{i}\big) 
=\begin{cases}
 \tau(q_{i}, q_j)  \times \epsilon(q_j, a_j) & i=n-1 \mbox{ and } a_i \not = n\\
\tau(q_{i}, q_j)  \times \frac{\eta(q_i, q_j, a_i, a_j)}{\epsilon(q_i, a_i)} & i=n-1 \mbox{ and } a_i = n\\
 \frac{\eta(q_i, q_j, a_i, a_j)}{\epsilon(q_i, a_i)} & i<n-1 \mbox{ and }
a_i = n \\
\end{cases}
\label{cond}
\end{equation}
\begin{definition} \rm For any integer $k\geq 0$, the $\alpha_k$-HMM is an $\alpha$-HMM  where in the underlying PIG,  $\forall r\in S$, influence $(q, r) \in A$ holds for at most $k$ distinct states $q \in S$. 
\end{definition}

Therefore, the conventional hidden Markov model (HMM) is the $\alpha_0$-HMM. That is the \ahmm\, without influence edges in the underlying PIG.

It is not difficult to see that the joint probability of the stochastic processes on any walk $\rho$ can be computed with  (\ref{chainrule}) together with (\ref{ahmm}). We point out, however, 
formula (\ref{ahmm}) does not specify how the the right-hand-side probability of $X_j$ conditional on multiple random variables should be   
is computed. This is similar to the Bayesian network setting, different methods may be adopted. Since 
variables $X_i$, $i\in F_{\rho}(n)$, may be interdependent, 
one viable method is the Noisy-OR  \cite{Srinivas1993}. While it is a heuristic method, normalization with the partition function can be applied to ensure a well-defined probability distribution, as used in clique factorization in Markov random fields and Gibbs fields \cite{Grimmett1973,Clifford1990}.

\subsection{RNA secondary structure}
\begin{figure}[ht]
\begin{center}
\fig[width=290pt]{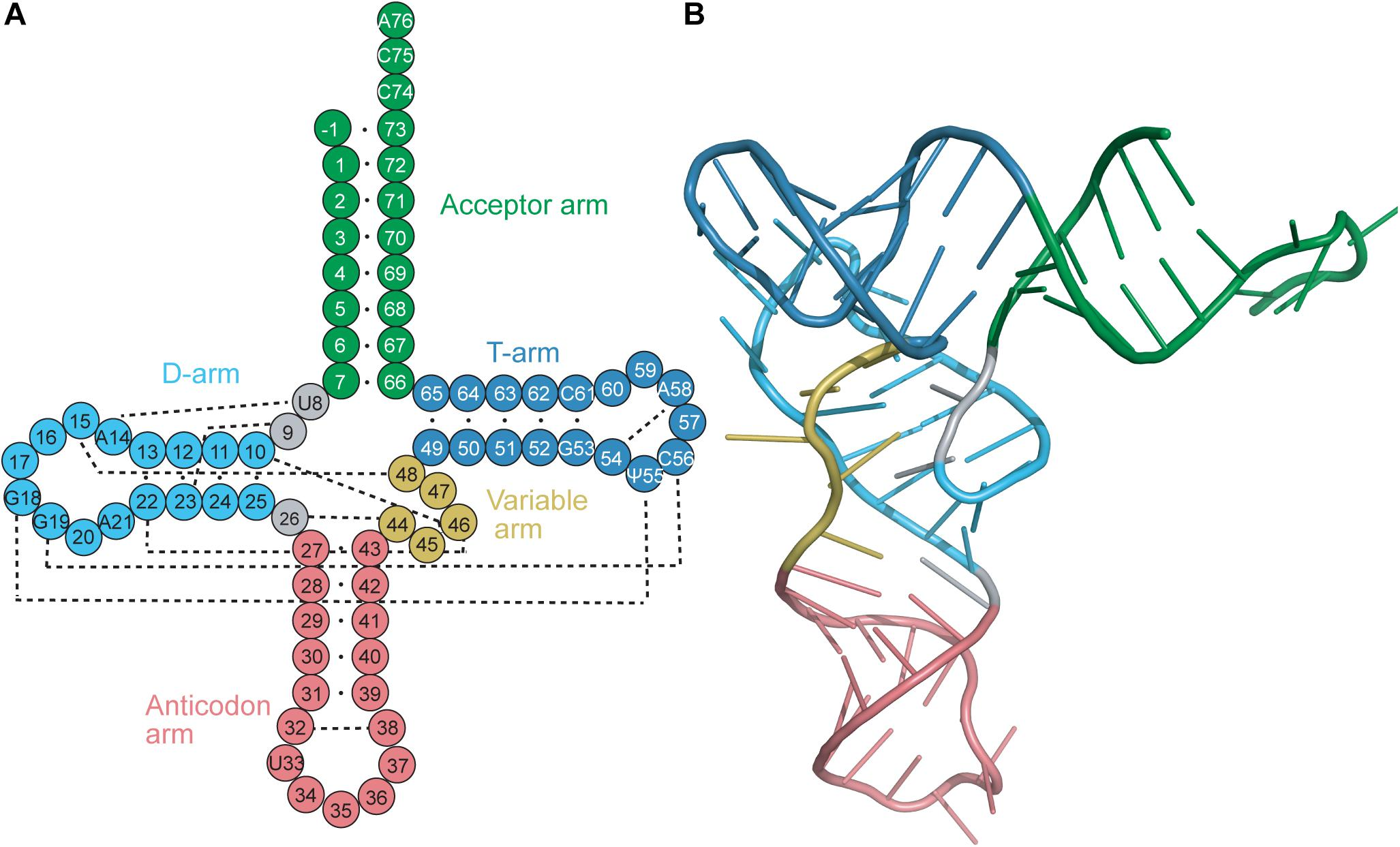}
\fig[width=140pt]{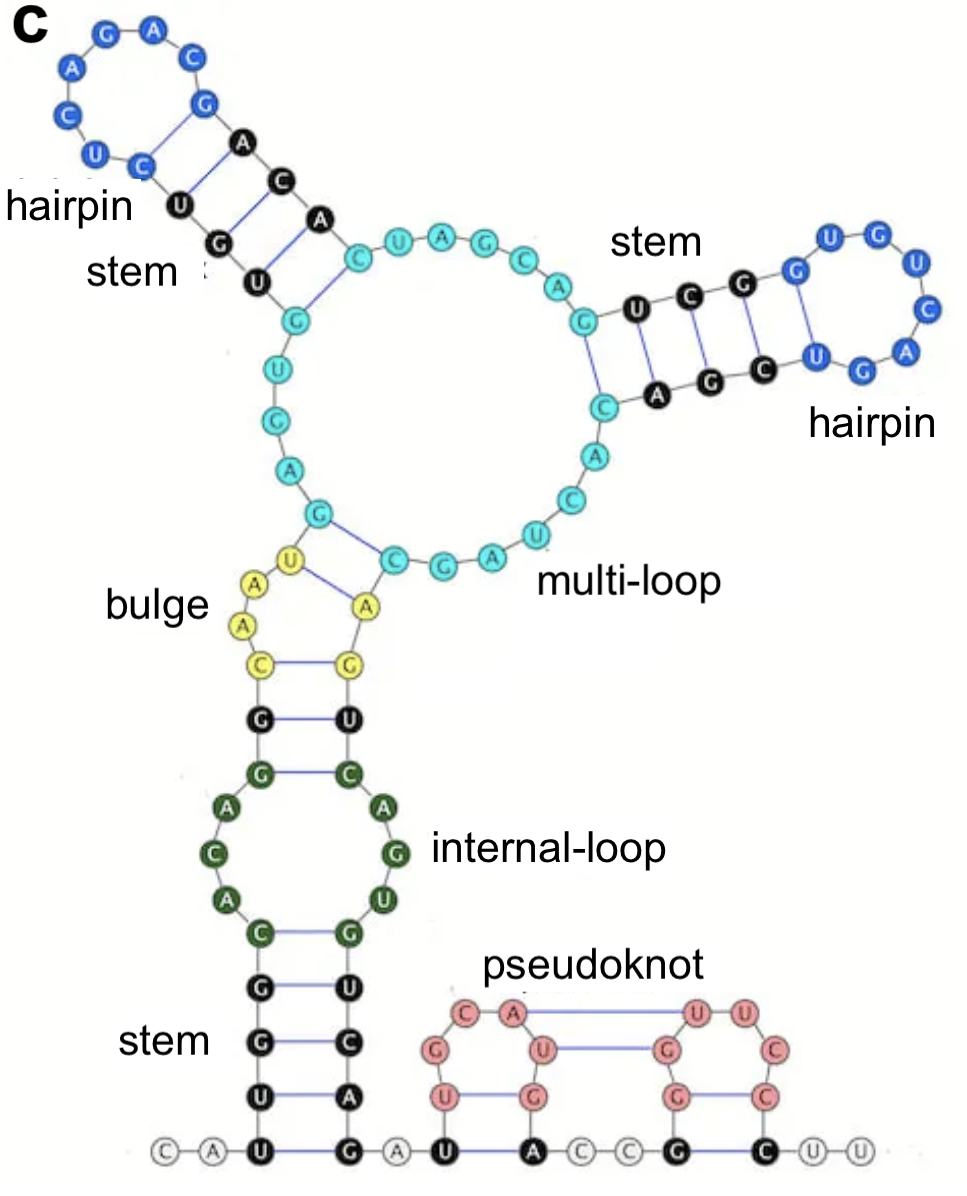}
 \end{center}
\caption{(A) Secondary structure of tRNA, where canonical pairs of nucleotides are connected with a single dot '.'  whereas non-canonical nucleotide interactions are represented by longer dotted lines. The four stems are colored code to correspond the four double helices in its 3D structure in (B).  (C) Various structural elements in RNA secondary structure, including stem, hairpin loop, bulge loop, internal loop, multi-loop, and pseudoknot.
}
\label{tRNA} 
\end{figure}

A ribonucleic acid (RNA) is a macro-molecule of nucleotides, each consisting of phosphate group, sugar, and aromatic base that is usually either adenine ({\tt A}), cytosine ({\tt C}), guanine ({\tt G}), or uracil ({\tt U}), with some exceptions and modifications. While phosphodiester bonds connect sugar and phosphate groups of nucleotides to make them form a linear sequence, nucleotides may also physically come close to each other through hydrogen bonds between bases, resulting in the {\it secondary structure}. Such base-base pairings are the canonical Watson-Crick pairs ({\tt A-U} and {\tt G-C}), and the wobble pair ({\tt G-U}). We note that such base pairs are different from much more evasive, non-canonical interactions between nucleotides that belong to the realm of RNA tertiary (3D) structure \cite{DoudnaEtAl1998,LeontisEtAl2008}. Figure~\ref{tRNA}~(A) shows the secondary structure of transfer RNA (tRNA) and its scaffolding role in RNA 3D structure in Figure~\ref{tRNA}~(B).

Two important components constitute the RNA secondary structure.
A {\it stem} typically consists of three or more stacking pairs of hydrogen-bonding nucleotides across two opposite strands in the RNA sequence. The involved 5' strand is called the {\it left arm} and the 3' strand is called the {\it right arm}. The biochemistry stability of a stem requires at least 3 stacked base pairs in general. The two strands forming the stems also need to be {\it spanned} by at least 4 other consecutive nucleotides to maintain the needed stability. An unpaired region of contiguous  nucleotides in the sequence is called a {\it loop}.  Loops connect stems in various ways resulting in different substructural elements of the secondary structure (see Figure~\ref{tRNA} (C)).

\subsection{Composing secondary structure}

Figure~\ref{tRNA} shows that an RNA secondary structure consists of various substructural elements. In this section, we describe these substructures as higher-order relationships over the nucleotides indexes $\{1, 2, \dots, n\}$ on the RNA molecule sequence of length $n$. Let $s$ be any contiguous strand of nucleotides on the molecule sequence. We represent $s$ with the closed set of consecutive integer indexes: $s =[s^{(5)} . . \, s^{(3)}]$, where 
$s^{(5)}$ and $s^{(3)}$, $s^{(5)} \leq s^{(3)}$, are  the indexes of the first and last nucleotide on the strand $s$, respectively.

\begin{definition}\rm A {\it stem} $(l, r)$ consists two strands $l$ and $r$ that satisfy (i) 
$2 \leq l\ptr - l\pfv = r\ptr - r\pfv$ and  (ii) $l\ptr \leq  r\pfv - 5$. A {\it loop} $s$ is a single strand, with $s\pfv \leq s\ptr$.
\end{definition}

Condition (i) requires that a stem has at least three base pairs, and (ii) ensures that the left arm $l$ and right arm $r$ of the stem are at least 4 nucleotide positions apart.

\begin{definition}
\rm
Let $Z$ be an RNA sequence of length $n$. The {\it secondary structure of $Z$} is a collection 
\[ SS_Z = \{ (l_i, r_i): (l_i, r_i) \mbox{ is a stem }, 1\leq i\leq k \}\bigcup \{s_i: s_i \mbox{ is a loop}, 1\leq i\leq h \}\]
 for some $k, h \geq 0$, such that 
  
\noindent
\dind  (1) $(l_i \cup r_i ) \bigcap (l_j \cup r_j ) = \emptyset$, for all $i\not =j$,  $1\leq i, j \leq k$;\\
 \dind (2) $s_i \bigcap s_j = \emptyset$, for all $i\not =j$, $1\leq i, j \leq h$;\\ 
 \dind (3) $(l_i, r_i) \bigcap s_j = \emptyset$ for all $1\leq i\leq k$ and $1\leq j\leq h$;\\
 \dind (4) $\big(\cup_{i=1}^k  (l_i, r_i) \big) \bigcup \big(\cup_{j=1}^h s_j\big) = [1..n]$
\end{definition}

 Conditions (1)-(3) require that no stems or loops overlap in their indexes and (4) ensures that all stems and loops make up exactly the indexes of the RNA sequence.

Based on the definitions, substructural elements in an RNA secondary structure can now be viewed as how stems and loops are positioned (according to indexes). 

\begin{figure}[ht]
\begin{center}
\fig[width=430pt]{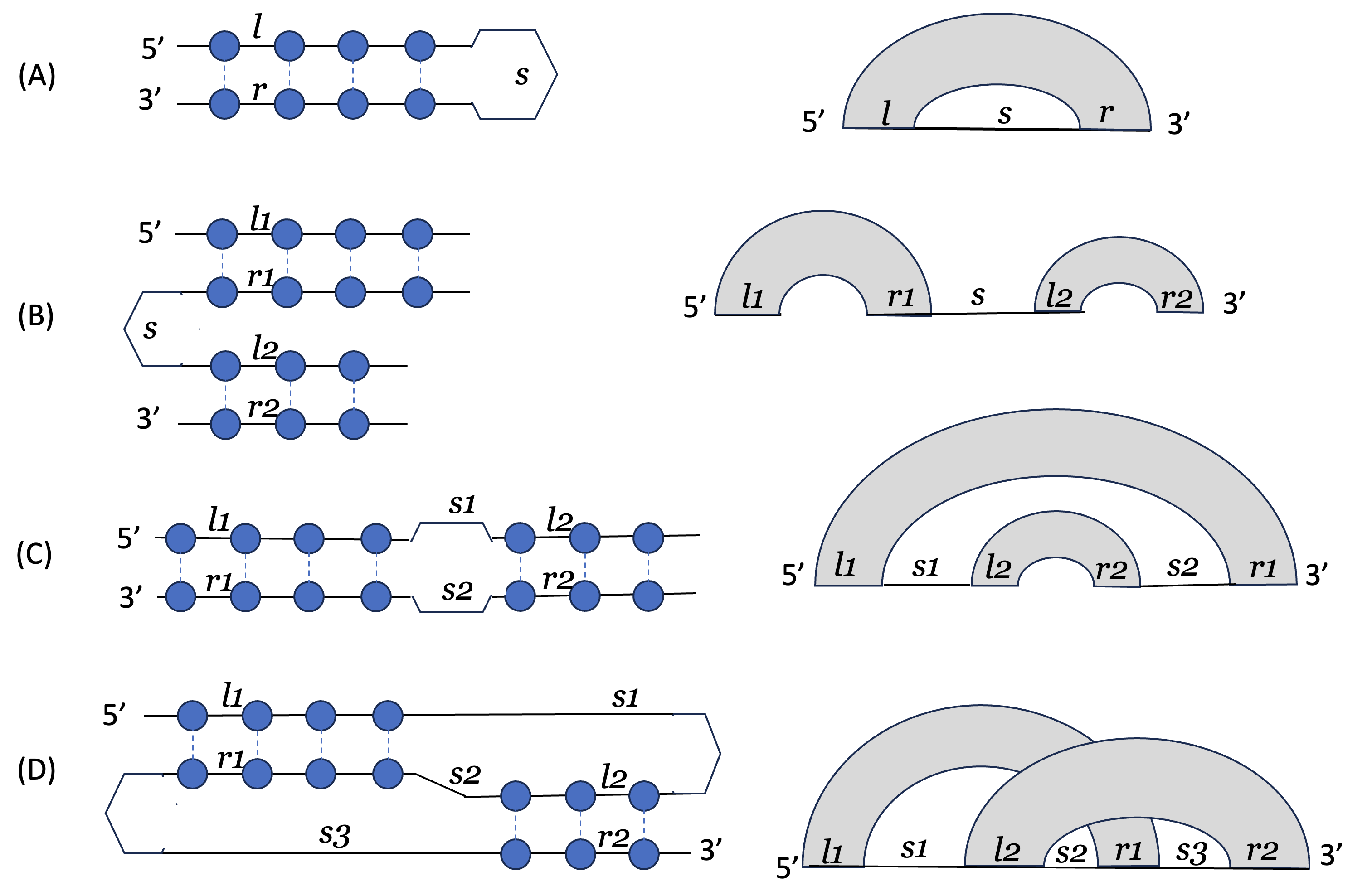}\,
 \end{center}
\caption{RNA composite stems.  (A) stem-loop, (B) parallel stems, (C) nested stems, and (D) crossing stems. The rules for forming composite stems can repeatedly be  applied to any unpaired loop region. 
}
\label{CompStems} 
\end{figure}

\begin{itemize}
\item[(A)] A {\it stem-loop} consists of a stem $(l, r)$ and a loop $s$ such that $l\ptr + 1 = s\pfv$ and 
$s\ptr +1 = r\pfv$ (Figure~\ref{CompStems}.A);
\item[(B)] Two {\it parallel stems} consist of two stems $(l_1, r_1)$, $(l_2, r_2)$, and a loop $s$ such that $r_1\ptr + 1 = s\pfv$ and $s\ptr+1 = l_2\pfv$ (Figure~\ref{CompStems}.B);
\item[(C)] Two {\it nested  stems} consist of two stems $(l_1, r_1)$, $(l_2, r_2)$, two loop $s_1$, and $s_2$, such that $l_1\ptr+1 = s_1\pfv$, $s_1\ptr +1 = l_2\pfv$, $r_2\ptr+1 = s_2\pfv$, and $s_2\ptr + 1 = r_1\pfv$ (Figure~\ref{CompStems}.C);
\item[(D)] Two {\it crossing stem} consists of two stems $(l_1, r_1)$, $(l_2, r_2)$, three loop $s_1$, $s_2$, and $s_3$, such that $l_1\ptr+1 = s_1\pfv$, $s_1\ptr +1 = l_2\pfv$, $l_2\ptr+1 = s_2\pfv$; $s_2\ptr+1 = r_1\pfv$, $r_1\ptr+1 = s_3\pfv$, and $s_3\ptr+1 = r_2\pfv$ (Figure~\ref{CompStems}.D);
\end{itemize}

Figure~\ref{CompStems} gives illustrations on these substructures formed by combinations of stems and loops with their indexes of nucleotides satisfying the requirements. These definitions can be repeatedly applied to replace any loop region eventually to form a whole secondary structure for RNAs. For example, the secondary structure of tRNA shown in Figure~\ref{tRNA} consists of three parallel stems enclosed by the 4th as nested stems. Secondary structures consisting of only parallel and nested stems are {\it context-free}; prediction algorithms based on both energy minimization and SCFG are of time complexity $O(n^3)$ on RNA sequences of length $n$. In contrast, crossing stems
(also called {\it pseudoknotted structures} and of the context-sensitive nature) are computationally intractable to predict \cite{LyngsoAndPedersen2000}. Pseudoknots may seem a slightly rare compared to other types of stems but they are still present in many newly discovered in a wide spectrum of RNAs and of various important functions.

\section{Modeling RNA secondary structure }

We now show that $\alpha_k$-HMM is powerful enough and suitable to model RNA secondary structure including pseudoknots and $k=1$ suffices.

\subsection{Modeling composite stems}

Figure~\ref{pig} gives a 4-state PIG of $\alpha_1$-HMM that model the RNA secondary structures including pseudoknots. 
Here are some details of how stems and thus various elements of the secondary structure can be  generated by walks on the PIG. 

\begin{figure}[ht]
\begin{center}
\fig[width=400pt]{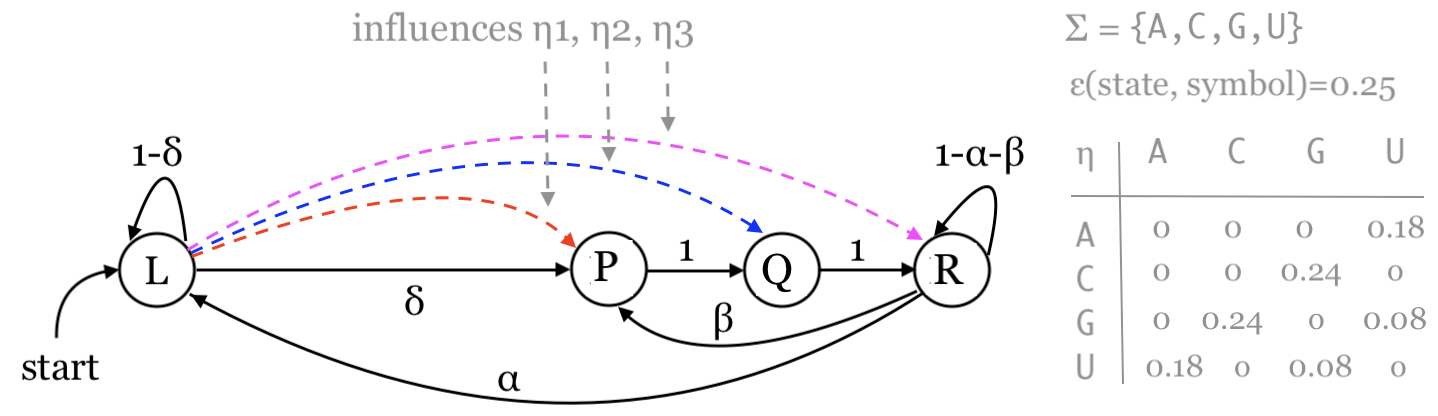}\,
 \end{center}
\caption{A simplified 4-state PIG for RNA secondary structure including pseudoknots. RNA sequences are assumed to be ``standard'' (with equal base composition for 4 nucleotides), i.e., every state has  the same emission probability $\epsilon = 1/4$ for the 4 nucleotides. Three (different) influence functions
$\eta_1, \eta_2,$ and $\eta_3$ are given as colored directed edges and their probability ``standard'' distributions  are summarized as the same distribution $\eta$ in the table. Transition function $\tau$ is given as the black directed edges. The values for probability parameters $\alpha$, $\beta$, and $\delta$ are to be determined in section 3.2. 
}
\label{pig} 
\end{figure} 

\begin{itemize}
\item Loops: Unpaired single strands, of length $\geq 1$, can be generated by state $L$, together with the self-loop transition $L\rightarrow L$, that  emits nucleotides on the strand.
\item  Stems: Stems can be generated by influences $\eta_1, \eta_2$, and $\eta_3$ of state $L$ onto
states $P, Q$, and $R$, respectively. Specifically, state $L$ generates nucleotides on the left arm of the stem and the influences together with state $P, Q$, and $R$, generates the base-paired nucleotides on the right arm. The self-loop $R \rightarrow R$ is allowed to generate the right arm of a matching length $\geq 3$ with the left arm (Figure Z-a). 

\item Parallel stems: For two or more stems positioned in parallel, once the first stem is generated, the second (and later stems) can be done in the same way simply by taking advantage of transition $R \rightarrow L$ to bring the stochastic process back to state $L$. In particular, the arms of two parallel stems are generated in the order of left1-right1-left2-right2 (Figure Z-b).

\item Nested stems: Likewise, two or more stems  positioned in a nesting fashion can be generated by repeating influences $\eta_1, \eta_2$, and $\eta_3$. However, instead of generating arms of the two parallel  stems in the order of left1-right1-left2-right2, the generating arms of two nested stems is done in the order of left1-left2-right2-right1. There are two possible scenarios between right2 arm and right1 arm: with and without unpaired nucleotides. The process takes the transition $R \rightarrow L$ for the former case and transition $R\rightarrow P$ for the latter (Figure Z-c).

\item Crossing stems: Similar to nested stems, but the arms of two crossing stems are generated in the order of left1-left2-right1-right2. Also similar to nested stems, there are two possible scenarios between right1 arm and right2 arm (Figure Z-d).
\end{itemize}

\subsection{Paramater estimation}

We now estimate probabilities for the parameters $\alpha, \beta$, and $\delta$ in Figure~\ref{pig}.

The stability of a stem, i.e., a stack of canonical base pairs, depends on the free energy contributed from the base pairs as a whole. The folding stability of a stem $\sigma$ requires the free energy threshold $\Delta E(\sigma)$ to reach a certain (low) level. Interestingly, it can be related to the ratio of probabilities between a structure model (stem) $\sigma$ and the null model (nucleotides independent otherwise) $I_{\sigma}$ (of the probability $\prod_{x\in \sigma} P(x)$). Specifically, 

\begin{equation}
\label{free-energy}
\frac{P(\sigma)}{P(I_\sigma)} = \frac{e^{-E(\sigma)}}{e^{-E(I_\sigma)}} = e^{-\big(E(\sigma) - E(I_\sigma)\big)} = e^{-\Delta E(\sigma)}
\end{equation}
where $\Delta E$ is a negative number, 
based on the Boltzmann distribution\footnote{Boltzmann distribution relates energy $E({\bf x})$ to probability $P({\bf x})$: $P({\bf x}) \propto e^{-E({\bf x})/{KT}}$ for Boltzmann constant $K$ and temperature $T$.} and 
\[ \frac{P(\sigma)}{P(I_\sigma)} = \prod_{x \diamond y \in \sigma} \frac{P(x, y)}{P(x)P(y)}\]
where $P(x, y)$ is the probability of base pair $x\diamond y$ between nucleotides $x$ and $y$ in $\sigma$ and $P(x)$ is the individual probability of $x$ in $\sigma$.

There are two sets of parameters in the \ahmm\, that need to be estimated. 

First, by equation~(\ref{free-energy}), for example, using free energy $\Delta E \approx -4.5 {\rm Kcal/mol}$ for stable stems, $e^{-\Delta E} \approx 90.017$ would be the probability ratio threshold to distinguish a stem stabled by its base pairs from its otherwise independent nucleotides. However, the \ahmm\,  in Figure~\ref{pig} involves transition probabilities other than emission probabilities $P(x)$ and influence probabilities $P(x, y)$, for the model to distinguish a stable stem from independent nucleotides, contributions from transitions need to be factored in as well. This leads to the following formula computing the odds:
\begin{equation}
\label{odds}
{\cal R} = \frac{P(\sigma)}{P(I_\sigma)}  \times f \times (\Delta_f)^{|\sigma|-3}
\end{equation}
where $f$ is the {\it transition factor}, together with $\Delta_f$ 
(the {\it incremental transition factor}),  is the ratio of the compound transition probability of the stem $\sigma$ against the compound transition probability of otherwise independent nucleotides in $\sigma$. In this work, we set the condition to distinguish a stable stem with ${\cal R} > 1$, or $\log {\cal R} > 0$. According to equations~(\ref{free-energy}) and (\ref{odds}), this leads to 
\[ 
f \times (\Delta_f)^{|\sigma|-3} >{e^{\Delta E(\sigma)}} 
\]

Therefore, given any statistically estimated free energy $\Delta E$ for stable for stable stems, $f$ and $\Delta_f$ can also be estimated, which make it possible to compute parameters $\alpha, \delta$, and $\beta$ for the \ahmm. There are two cases of stable stems to consider:
\begin{itemize}
\item  A standalone stem, of three consecutive base pairs, with the transition pattern $L\rightarrow P\rightarrow Q \rightarrow R \rightarrow L$ against $L\rightarrow L\rightarrow L\rightarrow L\rightarrow L$, 
with the smallest transition factor  $f_1 > \frac{1}{e^{-\Delta E}}$;

\item Two consecutive stems with transition patterns
$L\rightarrow P\rightarrow Q \rightarrow R \rightarrow P\rightarrow Q \rightarrow R \rightarrow L$ against $L\rightarrow P\rightarrow Q \rightarrow R \rightarrow L\rightarrow L\rightarrow L\rightarrow  L$, with  the smallest transition factor  $f_2 > \frac{1}{e^{-\Delta E}}$;

\end{itemize}
In both cases, more canonical base pairs can be added to the formed, stable stem, through self-loop transition $R\rightarrow R$, with the incremental transition factor $\Delta_f = \frac{1-\alpha - \beta}{1-\delta}$. 

These requirements together lead to

\[ \begin{cases}
f_1 = \frac{\delta \alpha}{(1-\delta)^4} > \frac{1}{e^{-\Delta E}} 
& (1) \\
f_2 =  \frac{\beta\alpha}{\alpha(1-\delta)^3}  > \frac{1}{e^{-\Delta E}} 
& (2) 
\end{cases}
\]

From (1) and (2), we select $f_2 = f_1$. Then $\beta = \frac{\delta \alpha}{1-\delta}$. Upon choose $\Delta E = -4.5{\rm Kcal/mol}$, for example, we may identify $\frac{1}{e^{-\Delta E} }\approx 0.0111$ and derive  $\delta = 0.05$, $\alpha = 0.2$, and $\beta = \frac{0.01}{0.95} \approx 0.0105 $, we obtain 
$f_1 = f_2= 0.0122 > \frac{1}{e^{-\Delta E} }$. In addition, $\Delta_f = \frac{1-\alpha - \beta}{1-\delta} = \frac{0.75}{0.95} \approx 0.7895$. 
 
 \vs
 Second, we estimate base pair probability $P(x, y)$ for $x, y \in \{{\tt A}, {\tt C}, {\tt G}, {\tt U}\}$ and single nucleotide probability $P(x)$. According to our analyses in the previous paragraphs, the base pair probability $P(x, y)$ actually refers to the probability of a base pair under the ``structured model'', which is the influence of state $L$ on state $P$, $Q$, or $R$. In particular, $P(x, y|L, S) = \eta(L, S, x, y)$, for every $S\in \{P, Q, R\}$. On the other hand, single nucleotide probability, i.e., $P(x|L) = e_L(x)$ refers to the probability of individual (not influencing nor influenced) nucleotides. 
 
 However,  base pair and single base probability distributions vary vastly across RNA sequences of different species and families. As an example Table~\ref{table-distributions}  shown the single base distribution $P(x)$ and base pair distributions $P(x, y)$ extracted from tRNA sequences whose secondary structures are predicted by software {\tt RNAfold} with 100\% accuracy. 
The score $S(x, y) = \frac{P(x, y)}{P(x)P(y)}$ is computed as the probability ratio between the two. With these data and $\Delta E = -4.5{\rm Kcal/mol}$, a stem $\sigma$ needs to have its compound score $\prod\limits_{x\diamond y \in \sigma} S(x, y)$  exceed the value $e^{-\Delta E} = 90.017$ to be stable. As a result, for stems consisting of 3 base pairs, only the one with all {\tt G}-{\tt C} pairs have its compound score $=134.45$ while all others fail to reach the level of $87$.

The above phenomenon is interesting since, with a slightly increase of $\Delta E$, say, to $-4.0{\rm Kcal/mol}$, the  threshold required for the compound score of a stable 3-bp stem is only 54.598, making it possible for a number of 3-bp (non-stable) stems to be predicted as being stable, incorrectly. On the other hand, with the original $\Delta E = -4.5{\rm Kcal/mol}$, the single base and base pair probability distribution extracted from a different set of RNA sequences yield a different score function $S(x, y)$ which is expected to predict well (only) on these data set with our \ahmm\, model. Therefore, the designed \ahmm\, reveals an intimate quantitative relationship between a minimum free energy threshold for stable stems and the base composition of the RNA sequences. 
 
 \begin{table}[ht]
\begin{center}
\begin{tabular}{|c||c|}
 \hline
 & $P(x)$\\
 \hline
 {\tt A} & 0.22\\\hline
 {\tt C} & 0.28 \\\hline
 {\tt G} & 0.29\\\hline
 {\tt U} & 0.21\\\hline
 \end{tabular}
 \quad
\begin{tabular}{ |c|| c| c | c|c|}
\hline
 $P(x, y)$ &  \,{\tt A}\,  & \,{\tt C} \,  & \,{\tt G} \, & \,{\tt U} \\ 
 \hline
 \hline
 {\tt A}  & 0 &  0 & 0  & 0.125  \\  
 \hline
 {\tt C}  & 0 &  0 & 0.267& 0  \\
 \hline
 {\tt G}  & 0 &  0.416 & 0 & 0.029  \\
 \hline
 {\tt U}  & 0.142 & 0  & 0.021  &  \\
 \hline
 \end{tabular}
 
 \vs
 \begin{tabular}{ |c|| c| c | c|c|}
\hline
 $S(x, y)$ &  \,{\tt A}\,  & \,{\tt C} \,  & \,{\tt G} \, & \,{\tt U} \\ 
 \hline
 \hline
 {\tt A}  & 0 &  0 & 0  & 2.706   \\  
 \hline
 {\tt C}  & 0 &  0 & 3.288& 0   \\
 \hline
 {\tt G}  & 0 &  5.123 & 0 & 0.476   \\
 \hline
 {\tt U}  & 3.074 & 0  & 0.345 &  0 \\
 \hline
 \end{tabular}
 \caption{Single nucleotide probability distribution $P(x)$ (left table) and base pair probability distribution (middle table) obtained from 150 tRNA sequences on which RNAfold predicts with 100\% accuracy. They are used to calculate {\it standard} odds function $S(x,y) = \frac{P(x,y)}{P(x)P(y)}$ (bottom table) for base pairing in our $\alpha$-HMM.}
 \label{table-distributions}
 \end{center}
 \end{table}

\section{Decoding algorithm for structure prediction }
 
The task of second structure prediction is to identify a most plausible collection of stable (non-conflicting) stems from the underlying linear sequence. Since the stems are modeled with influences, the structure prediction is equivalent to decode a most probable walk with influences on the PIG. We describe an optimal decoding algorithm for $\alpha_1$-HMM in this section.

The decoding task is to find an walk $\rho^*$ that maximizes the joint probability of the involved states and influences with the observation sequence ${\bf a} = a_1\dots a_n$. By equations~(\ref{chainrule}) and (\ref{ahmm}), 
\begin{equation}
\label{max-walk} \rho^* = \arg \max_{\rho= \lb q_1, k_1\rb  \dots \lb q_n, k_n\rb} P\big(\lb q_1, k_1\rb, a_1\big) \prod_{j=2}^{n} P\big (\lb q_j, k_j\rb, a_j \big | 
\cup_{i \in F_\rho(j)} \lb q_i, k_i\rb, a_i\big)
\end{equation}

where the individual conditional probability is computed with equation~(\ref{cond}). Since our model for RNA sequences is an $\alpha_1$-HMM, $|F_\rho(j)| \leq 2$, and  $j-1 \in F_\rho(j)$. Therefore, the conditional probability, for any $j\geq 2$, 

$P\big (\lb q_j, k_j\rb, a_j \big | 
\cup_{i \in F_\rho(j)} \lb q_i, k_i\rb, a_i\big)$ 
\begin{equation}
\label{max-walk-cond}
  = \begin{cases} 
   \tau(q_{j-1}, q_j) \epsilon(q_j, a_j) & \mbox{ if } |F_\rho(j)|=1 \mbox{ and } k_{j-1}\not =j\\
   \tau(q_{j-1}, q_j) \frac{\eta(q_{j-1}, q_j, a_{j-1}, a_j)}{\epsilon(q_{j-1}, a_{j-1})} & \mbox{ if } |F_\rho(j)| =1  \mbox{ and } k_{j-1}=j\\
 \frac{\eta(q_{l}, q_j, a_{l}, a_j)}{\epsilon(q_{l}, a_{l})} & \mbox{ if } |F_\rho(j)| = 2, l\in F_\rho(j),  l<j-1, \mbox{ and } a_l=j \\
\end{cases}
\end{equation}

Decoding with equation~(\ref{max-walk}) needs to compute probability $P(\rho, {\bf a})$ for every potential walk $\rho$ on the observation $\bf a$. Therefore, let $r\in S$ and $l<j \leq n$. We introduce  
recursive function $m(j, r, l)$ to be the maximum probability of a walk arriving at state $r$ in the $j$th step that is influenced by step $l$. Then by the chain-rule (\ref{chainrule}), equation~(\ref{max-walk}), and (\ref{max-walk-cond}), function $m$ can be recursively defined with
\begin{equation}
\label{DP}
 m(j, r, l) = \max_{s \in S, 1\leq k < j-1} \big \{ m(j-1, s, k) \times P\big (\lb q_j, k_j\rb, a_j \big | 
\cup_{i \in \{l, j-1\}} \lb q_i, k_i\rb, a_i\big)\big\}
\end{equation}

with base case: $m(1, r, l) = \epsilon(s_0, a_1)$ if and only if $r=s_0$, the starting state  $s_0$, and $l=0$.

The computation of $m(j, r, l)$ with equation (\ref{DP}) needs to follow the rules for influences in Definition 1 for PIG and in Definition 3 for walks. Specifically, let $m(j-1, s, k)$ be such upon which 
maximum probability of $m(j, r, l)$ is established and let $\rho^{(j-1)} = \lb p_1, h_1\rb \dots \lb p_{j-1}, h_{j-1}\rb$ be the walk (up to step $j-1$) corresponding to the probability value $m(j-1, s, k)$. Then the walk $\rho^{(j)} = \lb q_1, k_1\rb \dots \lb q_j, k_j\rb $ (up to step $j$) that corresponds to probability value $m(j, r, l)$ 
should satisfy the following validities:

(1) $\forall i, 1\leq i \leq j-1$, $q_i = p_i$;


(2) $\forall i, 1\leq i \leq j-1$, if $i\not = l$, then $k_i=h_i$;

(3) $h_l = 0$ and $k_l = j$;

(4) if $r$ is affiliated with $s$ and $k\not = 0$, then $k=l+1$;

\begin{figure}
\begin{center}
\fig[width=250pt]{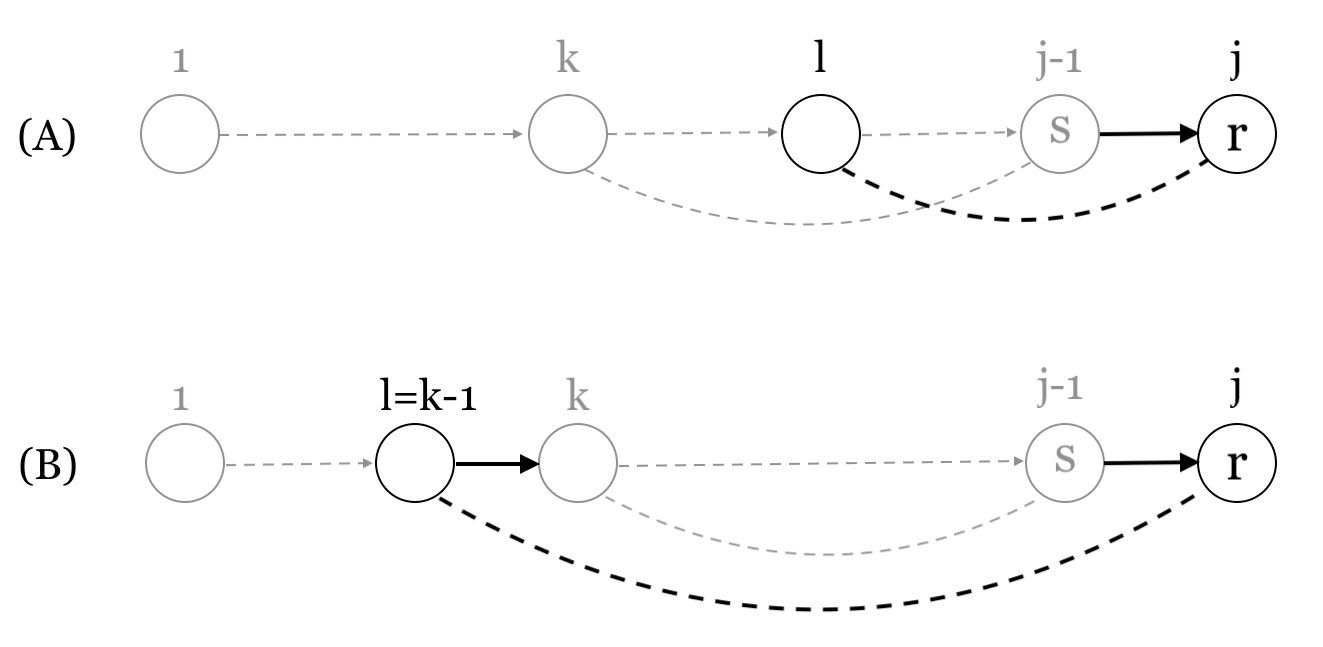}
 \end{center}
\caption{Illustration on walk $\rho^{(j)}$ up to step $j$ is built upon established walk $\rho^{(j-1)}$ up to step $j-1$ (gray drawings). (A) Potential new influence to step $j$ from step $l$, as long as the latter has not been a influencer. (B) When state $r$ is affiliated with its predecessor state $s$, the influence from step $l$ to step $j$ needs to be established upon influence from step $l+1$ to step $j-1$, should $j-1$ is influenced in walk $\rho^{(j-1)}$.}
\label{dp-figure} 
\end{figure}

\vs
\noindent
In other words,  walk $\rho^{(j)}$ is built (incrementally) from established walk $\rho^{(j-1)}$ with potential new influence from step $l$ to the new step $j$, while keeping all the states from walk $\rho^{(j-1)}$. It requires that walk $\rho^{(j)}$ does not create a new influence on any step before the new step $j$ neither that step $l$ has been an influencer in walk $\rho^{(j-1)}$. In addition, if state $r$ of the new step $j$ is affiliated with its predecessor state $s$ of step $j-1$, then the two influences to from step $l$ to step $j$ and from step $k$ to step $j-1$ are in the nested fashion, i.e., $k=l+1$.  Figure~\ref{dp-figure} illustrates these scenarios.

Therefore, the function $m$ can be computed with a dynamic programming algorithm that establishes a table for function $m$. The information of the corresponding walk for each entry $m(j, r, l)$ can be stored, incrementally updated, and looked up by the computation of subsequent entries. The total time complexity of the algorithm is $O(n^3 |S|^2)$, where $n$ is the length $|{\bf a}|$ of observation ${\bf a}$, and $|S|$ is the size of model, which is usually bounded by a very small constant.

It is not difficult to see that all influences on the optimal walk predicted can be simply traced back, which yield all the base pairs that constitutes the secondary structure.
 

\section{Conclusion}

We have introduced probabilistic graphical model \ahmm, as an extension to the HMM, capable of modeling  higher-order structures possessed by the modeled stochastic process. In particular, the \ahmm\, offers a convenient mean define correlations between present events and distant, historical ones through the notion of influence. This results in a specific framework to account for the canonical pairing between two, possibly distant nucleotides on an RNA sequence. We have shown a succinct \ahmm, along with estimated parameters, to model not only stem-loops and multi-loops but also pseudoknots of RNA secondary structures. We also present a general dynamic programming based decoding algorithm that can be tailored to the model for the prediction of RNA secondary structure including pseudoknot. The algorithm is very efficient with time complexity $O(n^3)$ on the input RNA sequence consisting of $n$ nucleotides.

The \ahmm\, includes the expressibility of the SCFG and beyond. Specifically, context-free rewriting rules in the SCFG can be emulated by transition and influence edges in the \ahmm\, graph. Therefore, SCFG-based RNA secondary structure prediction algorithms \cite{RivasEtAl2011,AndersonEtAl2012,DingEtAl2014} can be recaptured by our work to yield the same performance if not better. In particular, since the $\alpha$-HMM can flexibly apply restrictions on how one event may influence another in a stochastic process,  the given \ahmm\, in Figure~\ref{pig} can be extended to include stacked base pairs for a more accurate account for free energy. We leave the implementation of the decoding algorithm and the secondary structure prediction tests on various families of RNA secondary structure to another forthcoming publication of ours. 

\newpage
\bibliographystyle{acm}
\bibliography{AHMMRNA-paper-2024-1}

\end{document}